# COMPETITION OF BREAKUP AND DISSIPATIVE PROCESSES IN PERIPHERAL COLLISIONS AT FERMI ENERGIES

T.I. Mikhailova[1], A.G. Artyukh[1], M. Colonna[2], M. Di Toro[2], B. Erdemchimeg[1,3], G. Kaminski[1,4], I.N. Mikhailov[1], Yu.M. Sereda[1,5], H.H. Wolter[6]

[1] *Joint Institute for Nuclear Research, 141980, Dubna, Russia*
[2] *Lab. Naz. del Sud, INFN, I-95123 Catania, Italy*
[3] *Mongolian National University, Nuclear Research Center, ROV 46A/305, Ulan-Baator, Mongolia*
[4] *Institute of Nuclear Physics PAN, Radzikowskiego 152, 31-342 Krakow, Poland*
[5] *Institute for Nuclear Research NAS, 252650, prospect Nauki 47, Kyiv-22, Ukraine*
[6] *Faculty of Physics, University of Munich, 85748 Garching, Germany*

Heavy ion collisions in the Fermi energy regime may simultaneously show features of direct and dissipative processes. To investigate this behavior in detail, we study isotope and velocity distributions of projectile-like fragments in the reactions $^{18}$O (35MeV/A) + $^{9}$Be($^{181}$Ta) at forward angles. We decompose the experimental velocity distributions empirically into two contributions: a direct, 'breakup' component centered at beam velocity and a dissipative component at lower velocities leading to a tail of the velocity distributions. The direct component is interpreted in the Goldhaber model, and the widths of the velocity distributions are extracted. The dissipative component is then successfully described by transport calculations. The ratio of the yields of the direct and the dissipative contributions can be understood from the behavior of the deflection functions. The isotope distributions of the dissipative component agree qualitatively with the data, but the modification due to secondary de-excitation needs to be considered. We conclude, that such reactions are of interest to study the equilibration mechanism in heavy ion collisions

## 1. Introduction

Peripheral collisions in the Fermi energy regime have gained much attention recently, because they appear to be a powerful method to produce new isotopes away from the stability line [1]. In this energy region the velocities of the incident motion of ions and of the Fermi motion are of the same magnitude. Therefore there is a competi-tion between processes that are mediated by direct nucleon-nucleon interactions with those, where many nucleons are involved, and which lead to dissipative processes. This can be seen in the velocity distributions of light fragments which have similar features at low and at relativistic energies [2-5]. Different models are conventionally used to describe both types of processes.

The direct process can be seen as a breakup process, where a part of the projectile is abraded in the collision with the target, which is implied in abrasion-ablation model. The shape of velocity distributions of light fragments produced in these processes can be interpreted in the Goldhaber model, in which the reaction between the two ions is described as independent removal of nucleons from the projectile [6].

The dissipative, or deep-inelastic (DIC), collision is characterized by an essentially binary collision process but with a substantial exchange of mass and energy. For a review of theoretical approaches that have been used to model it, see ref.[7] and refs therein. These approaches range from classical friction models with empirical or microscopic transport coefficients to the two-center shell model. More recently they have also been interpreted in the framework of semi classical transport theories of the BNV (Boltzmann-Nordheim-Vlasov) [8,9] and the QMD (Quantum Molecular Dynamics) types [10]. These models include the mean field dynamics and nucleon-nucleon collisions on an equal footing. Earlier applications of transport theory to deep-inelastic collisions showed qualitative agreement with experiment [4,11]. However, experimental data were rather global, and couldn't get insight at specific features of the process.

In this work, we investigate velocity and isotope distributions of projectile-like fragments (PLF) obtained in the reactions $^{18}$O on $^{181}$Ta and $^{9}$Be at 35 MeV per nucleon [10,12]. The experiment was performed with the use of magnetic spectrometer COMBAS at FLNR in Dubna (see the report to this conference of G. Kaminski). In the experiment the velocity distributions of fragments were measured at forward angles ($\theta$<2.5°), for a wide range of PLF's from He to nuclei slightly heavier than the projectile. An interpretation with QMD transport calculations [10] failed to describe the complete velocity distribution.

In an earlier work [8] we have proposed to decompose the velocity spectra into two components on empirical grounds, a Gaussian distribution centered at beam velocity, and a dissipative contribution, described in a transport model. In this work we present a more quantitative study of this mechanism. We investigate the systematic behavior of the width of the direct and the relative magnitude of the direct and the dissipative components. We show that there is connection between the asymmetry of the velocity distribution of light fragments and the shape of the deflection curves.

## 2. Theoretical description

Here we apply the Boltzmann-Nordheim-Vlasov (BNV) transport approach [13] to describe DIC in the Fermi energy regime. This approach has been used extensively in the last decades to describe higher energy heavy ion collisions. The BNV approach describes the time evolution of nucleon one-body density function in phase space $f(\vec{r},\vec{p},t)$ under influence of a mean field $U(f)$

$$\frac{\partial f}{\partial t} + \frac{\vec{p}}{m}\nabla_r f - \vec{\nabla}_r U \vec{\nabla}_p f = I_{cls}[f,\sigma]. \qquad (1)$$

The potential $U(f)$ is the sum of Coulomb and nuclear potentials, the later one is defined by effective Skyrme forces. The collision term $I_{cls}$ takes into account the Pauli blocking. The integro-differential equation (1) is solved using the test particle method as described in ref. [13].

The calculations proceed until a freeze-out time $t_f$, at which the number of fragments is stable, and where their properties are determined. A fragment is defined by a surface chosen by the condition that the density $\rho$ falls to 0.03 $\rho_0$, where $\rho_0$ is the saturation density. To compare the results with experiments in the restricted angular acceptance, Coulomb trajectories for each fragment are attached at freeze-out [8]. This also allows us to calculate average deflection curves, which are shown in Fig. 1. The band around angle zero is the angular acceptance of the experiment. For the heavier target ($^{181}Ta$) only a small range of dissipative trajectories contribute to the forward angle cross section, while for the light target ($^9Be$) a large range of impact parameters from Coulomb to grazing trajectories contributes to it. It will be seen that this is important for the shape of the velocity spectra.

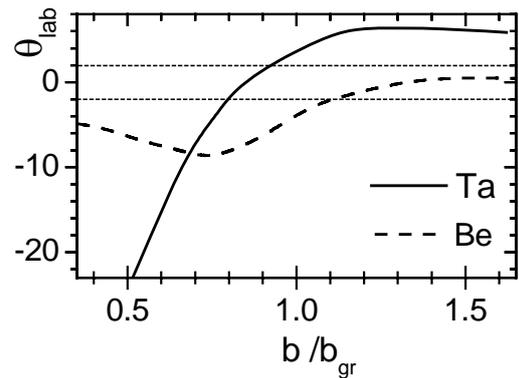

Fig. 1. Deflection curves as a function of impact parameter , solid line - the reaction $^{18}O + ^{181}Ta$, dashed one – $^{18}O + ^9Be$, 35 Mev/nucl

## 3. Results of calculations and comparison to experiment

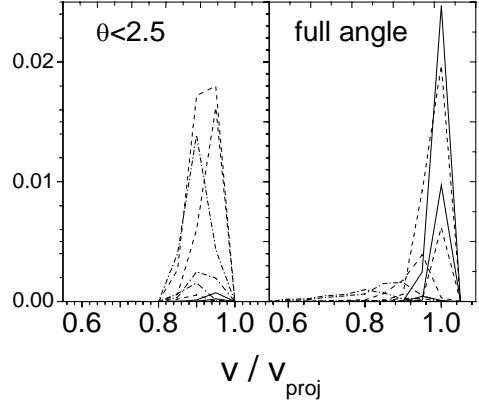

Fig. 2. Calculated velocity distributions for C, N and O isotopes in the reaction $^{18}O+^{181}Ta$, 35 AMeV with (left panel) and without (right panel) angular restrictions.

Here we present the results of calculations for the reactions $^{18}O +^{181}Ta$ and $^{18}O +^{9}Be$ at 35 AMeV, where we obtain velocity spectra and isotope distributions of the PLF's. In Fig. 2 we show the calculated velocity distributions for the reaction on the heavy target for different isotopes of oxygen (solid), nitrogen (dashed) and carbon (dashed-dot line). The calculated velocity distributions emitted at forward angles (θ<2.5˚) are always peaked at values of the relative velocity $v/v_{proj}$ less than one, moving to lower velocities when the mass of PLF decreases. Without the restriction of the emission angle, the fragment velocity can reach the value of projectile velocity, because then we include non-dissipative trajectories. One can also notice that the yield of oxygen isotopes in the case of the angular restrictions is the minimal, though it becomes the most abundant element when we include the grazing angle region in our calculations. Contrary to our calculations of the dissipative component of the reaction, the main feature of the experimental velocity distributions for all isotopes not heavier than the projectile is that their maxima are always close to the velocity of the projectile. The velocity distributions are very asymmetric and have slopes to the left with the pronounced shoulder, see Fig. 3, solid curve. This shoulder is more evident in the case of light target (i.e. the case of the inverse kinematics). As dis-cussed, we assume that there are two components in the experimental distributions. We separate the contribution of the DIC and BU processes in the experimental veloc-ity by a simple procedure, which is demonstrated in Fig.3. We fit the velocity distribution to the right of the maximum with a Gaussian and extract its width. We subtract this part from the total yield to obtain the dissipative component (Fig. 3, dashed curve). The maximum of this component is compared to the maximum on the calculated velocity distribution in Fig.4. There is a good correspondence between the two. Thus we find, that the dissipative component is well described by the transport calculations.

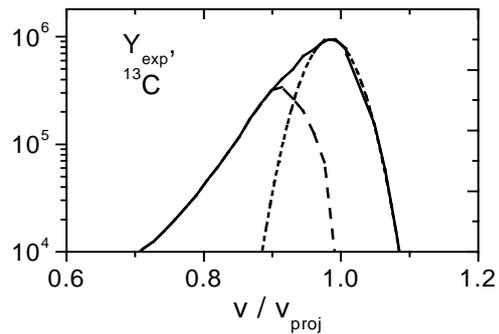

Fig 3. Decomposition of a typical experimental velocity distribution (solid curve) for the reaction of Fig 1, into a breakup (dotted line), and a DIC (dashed line) component.

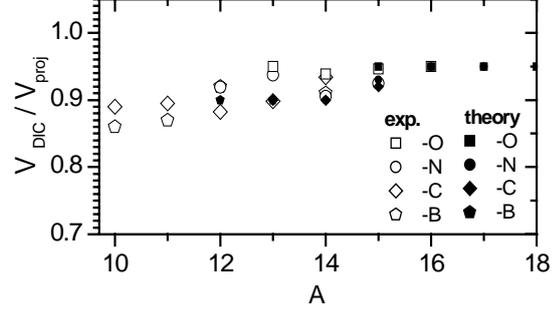

Fig. 4. Ratio of velocities at maximum to the projectile velocity for the dissipative component as a function of mass number A for the reaction of Fig.1. Full symbols – theory, open – experiment.

For the BU contribution we compare the mass dependence of the extracted widths of the different isotopes to the expression given in the Goldhaber model

$$\sigma_G^2 = \sigma_0^2 \frac{A_F(A_P - A_F)}{A_P - 1}, \quad \sigma_0 \approx 90 \text{ MeV}/c$$

Here $A_P$ is the mass number of projectile, $A_F$ is the mass number of the fragment and $\sigma_0$ is the normalization constant.

We find that the mass dependence of the width agrees well with the above expression. The normalization factor is about $\sigma_0^* = 60 \text{ MeV}/c$ and is the same for both reactions, which is less then the value of ref. [6], obtained from the assumption of independent removal of nucleons from a Fermi distribution at saturation density. This is consistent with the findings of other authors [3,5].

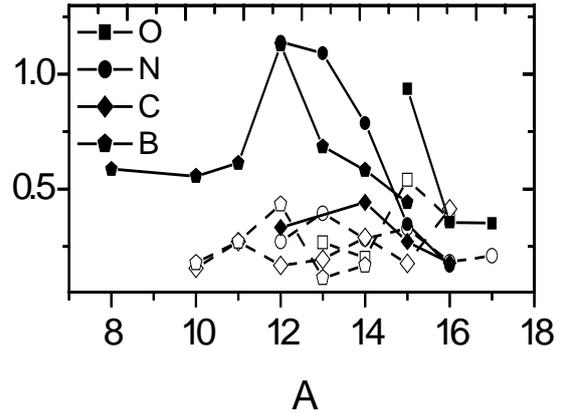

Fig. 5: Ratio of total yields DIC / BU components extracted from experimental data, full symbols – $^9$Be target, open symbols – $^{181}$Ta target

The ratio of the two modes, DIC and BU, thus determined for the two reactions is shown in Fig. 5. It is seen that dissi-pative processes constitutes an appreciable part of the cross section in the case of the light target $^9$Be, but is much smaller in the case of the heavy target $^{181}$Ta. This can be understood from the deflection curves for the two reactions in Fig. 1. For the experimental angular restriction of θ <2.5° only a narrow range of impact parameters contributes to the yield at forward angle for the $^{181}$Ta target, while for $^9$Be target, a wide range of impact parameters, including very distant trajectories, contributes to fragments moving in the forward direction. Thus the asymmetry of velocity spectra in these reactions depends on the colliding system, and is more pronounced in the case of inverse geometry.

We have also compared the isotope yields for the dissipative component from the data with the transport calculations. In general there is a good correspondence in the shape, however, the yield of small PLF's is too low [8]. However, in the calculation we deal with the primary, excited fragments at freeze-out time, while the experiment gives the final cold fragments. The de-excitation and evaporation of the primary fragments will be taken into account in the further work.

**Conclusion**

The Fermi energy regime is marked by a competition between direct and dissipative processes in heavy ion collisions. Here we have investigated these mechanisms in more detail by studying in particular forward angle scattering, where the direct process is expected to carry a bigger weight. To do so we have analysed recent experiments, obtained at the FLNR in Dubna of $^{18}$O collisions on various light and heavy targets, where isotopic distributions of projectile-like fragments as well as their velocity spectra had been measured.

We find that a good description of the data is obtained in terms of two components in the velocity spectra. One (dominant) component is centered around beam velocities with a Gaussian shape; the other shows a substantial energy loss with a long tail on the low-energy side. The first component is seen as a fast break-up process and interpreted in the Goldhaber model. The extracted widths follow the systematics of the Goldhaber model, the overall width constant is, however, lower than in the assumption of a free Fermi gas at saturation density, as was also found in other investigations.

The remaining dissipative contribution is described well by the transport calculations with respect to the shape of the velocity spectra and the mean dissipated energy. The isotope distributions of the dissipative component are also qualitatively reproduced, but are generally too narrow. We attribute this to the secondary decay of the calculated excited fragments, which was not yet included. We also investigate the ratio of the yield of the dissipative to the break-up component, which we can correlate to the behaviour of the deflection curves for the two reaction systems.

Our investigation shows that details of the dissipative component in heavy ion collisions at Fermi energies can be described by transport theory. A more microscopic description of the break-up component, however, remains to be developed. Generally we see that these reactions show a number of interesting phenomena in nuclear dynamics, and are in addition a useful tool to produce a large range of nuclei away from stability.

We would like to thank V.Baran, V.V. Pashkevich, F. Hanappe, V. Kuzmin, M. Zielinska-Pfabe for very helpful discussions, and V.Baran especially for the use of his codes.

REFERENCES

1. *Veselsky M*. et al, Heavy residues with A < 90 from the asymmetric reaction of 20 AMeV $^{124}$Sn + $^{27}$Al as a sensitive probe of the onset of multifragmentaion// Nucl. Phys. –2003– Vol A724. P. 431-454
2. *C.K. Gelbke et al,* Influence of intrinsic nucleon motion on energy spectra and angular distributions for $^{16}$O-induced reactions at 20 MeV/A, Phys. Lett –1977- Vol B70 – P. 415-417.
3. *Lahmer W. et al.* Transfer and fragmentation reactions of $^{14}$N at 60 MeV/u // Z. Phys. –1990- V. A337. - P. 425-437.; *Bacri Ch. O.* measurements in the beam direction of the $^{40}$Ar projectile fragmentation at 44 MeV/A // Nuc.Phys. – 1993 - Vol A555 – P.477-498
4. *Borrel* V. *et al* Peripheral Ar induced reactions at 44 MeV/u –similarities and deviations with respect to a high energy fragmentation process // Z. Phys –1983- Vol A73 –P. 191-197
5. *Notani M*. et. al. Projectile fragmentation reactions and production of nuclei near the neutron drip line// Phys. Rev.-2007 –Vol.C76 – P. 044605-1 – 044605-15.

6. *Goldhaber A.S.* Statistical model of fragmentation processes // Phys. Lett. – 1974 – Vol B53. - P. 306-308.

7. *U. Schröder and J.R. Huizinga*, Damped nuclear reactions, Treatise on Heavy-Ion Science Vol 2, ed. A.~Bromley, Plenum, New York, p. 113 (1984)8.

8. *Mikhailova T. et. al.* Investigation of dissipative collisions with transport models //Rom. Journ. Phys.-2007 –Vol. 52 – N8 – 10 – P 875-893

9. *Mikhailova* T. et al. Description of peripheral collisions at Fermi energies with transport models //Bull. of the Russian Academy of Sciences. Phys. – 2008 - Vol. 72 - N3 – P 363-368

10. *Artukh A.G. et al*. QMD approach in the descriptionof the $^{18}$O + $^{9}$Be and $^{18}$O + $^{181}$Ta reactions at $E_{proj}$ = 35 AMeV// Acta Physica Polonica. –2006 - V. 37 -. P. 1875-1892.

11. *M.F. Rive*t, et al., Dynamical aspects of violent collisions in Ar + Ag reactions at *E/A*=27 MeV, Phys. Lett. –1988- Vol B215 - P55

12. *A.G Artukh et al.*, Some regularities in the beam-direct production of isotopes with 2<*Z*<11 induced in reactions of $^{18}$O (35 A MeV) with Be and Ta, Nucl. Phys. –2002- Vol A701 – P 96

13. *G.F. Bertsch, S. Das Gupta*, A guide to microspcopic models for intermediate energy heavy-ion collisions, Phys. Rep. – 1988 – Vol 160 - P 189-260